# Enabling automated driving by ICT infrastructure: A reference architecture

Michael Buchholz[a]*, Jan Strohbeck[a], Anna-Maria Adaktylos[b], Friedrich Vogl[b], Gottfried Allmer[b], Sergio Cabrero Barros[c], Yassine Lassoued[c], Markus Wimmer[d], Birger Hätty[d], Guillemette Massot[e], Christophe Ponchel[e], Maxime Bretin[e], Vasilis Sourlas[f], Angelos Amditis[f]

[a] *Universität Ulm, Institut für Mess-, Regel- und Mikrotechnik, Albert-Einstein-Allee 41, D-89081 Ulm, Germany*
[b] *Autobahnen- und Schnellstraßen-Finanzierungs-Aktiengesellschafft (ASFINAG), Rotenturmstraße 5-9, Wien 1011, Austria*
[c] *IBM Ireland Limited, IBM House, Shelbourne Road, Ballsbridge Dublin 4, Ireland*
[d] *NOKIA Solutions and Networks GmbH, Werinherstraße 91, D-81541 München, Germany*
[e] *Airbus CyberSecurity SAS (AIRBUS), Boulevard Jean Moulin 1 ZAC de la clef Saint Pierre, Elancourt 78990, France*
[f] *Institute of Communication and Computer Systems (ICCS), Patission Str. 42, Athens 10682, Greece*

**Abstract**

Information and communication technology (ICT) is an enabler for establishing automated vehicles (AVs) in today's traffic systems. By providing complementary and/or redundant information via radio communication to the AV's perception by on-board sensors, higher levels of automated driving become more comfortable, safer, or even possible without interaction by the driver, especially in complex scenarios. Additionally, communication between vehicles and/or a central service can improve the efficiency of traffic flow. This paper presents a reference architecture for such an infrastructure-based support of AVs. The architecture combines innovative concepts and technologies from different technological fields like communication, IT environment and data flows, and cyber-security and privacy. Being the basis for the EU-funded project ICT4CART, exemplary implementations of this architecture will show its power for a variety of use cases on highways and in urban areas in test sites in Austria, Germany, and Italy, including cross-border interoperability.

* Corresponding author. Tel.: +49-731-50-27003;
  *E-mail address:* michael.buchholz@uni-ulm.de

# 1. Introduction

The automation of road transport is one of the major trends in mobility. Therefore, numerous initiatives, research projects and development initiatives are working on tomorrow's vehicles with automation level 3 and higher, e.g. [1], [2], [3]. However, the ability of a single vehicle to perceive its environment is limited by the available sensor technologies, leading to incomplete information of the automated vehicle's (AV's) environment, especially in complex scenarios. To overcome this issue, using information and communication technology (ICT) is an already established means to support AVs by external information, e.g., from other connected vehicles, from infrastructure like traffic lights or sensors, or from the cloud [4]. Therefore, ICT is an important enabler to establish automated driving, which will result in mixed traffic with manually driven vehicles for a long time. With the information provided by ICT, automated driving becomes more comfortable, safer, or even possible without interaction by the driver, especially in complex scenarios [5]. Additionally, communication between vehicles and/or a central service can improve the efficiency of traffic flow [5].

There are still many ICT-related challenges to overcome, in particular those related to the connectivity required for advanced levels of road vehicle automation and the architecture of such a connected ICT infrastructure. In addition, due to the external interfaces, cyber-security as well as privacy become inevitable to be investigated and integrated in such a system. One main challenge is to create ICT with a degree of harmonization between countries that will allow automated vehicles to traverse European borders freely. Harmonization of ICT functionalities and interfaces would also seem key for benchmarking and optimizing issues such as costs (investment, operation and maintenance) and requirements for latency, throughput, congestion strategies, data verification and data integrity.

The project "ICT Infrastructure for Connected and Automated Road Transport" (ICT4CART) [3] aims to develop and implement a respective ICT infrastructure to support and enable automated driving on higher levels. The project brings together key players from automotive, telecom, and IT industries to shape the ICT landscape for Connected and Automated Road Transport (CART) and to boost the EU competitiveness and innovation in this area. ICT4CART aims to provide concrete, evidence-based input feeding and standardisation processes (notably supporting interoperability and cyber security) and policy decisions (e.g. for spectrum policy). The project uses high-value use cases (urban and highway), to demonstrate and validate the harmonization concepts in real-life conditions at the test sites in Austria, Germany and Italy [6]. Significant effort will also be put on cross-border interoperability, setting up a separate test site at the Italian-Austrian border.

ICT4CART's ICT infrastructure considers the "Extended Vehicle" concept (ISO standards 20077 & 20078), where the horizontal integration of a cloud platform to enable this interconnected environment plays a central role and the cloud [7] and the free flow of data [8] initiatives launched by the EC are duly considered. ICT4CART also adopts the EC's common security and certificate policy[†] for all certificate authorities, which is originating from the work carried out in the C-ITS platform [9], [10]; and is based on Public Key Infrastructure (PKI) and the European C-ITS Trust model.

*1.1. Related work*

In the European ecosystem, several EU funded research projects enable ICT as a means to support AV's. Indicatively, 5G-MOBIX, 5GCroCo and 5G-CARMEN are three automotive projects which have been selected by the EC in response to the 5G-PPP ICT-18-2018 Call. The 5G Infrastructure Public Private Partnership (5G PPP) is a joint initiative between the European Commission and European ICT industry with the aim to deliver solutions, architectures, technologies and standards for the next generation communication infrastructures. 5G-MOBIX [11] will develop and test automated vehicle functionalities using 5G core technological innovations along multiple cross-border corridors (one between Spain and Portugal and one between Turkey and Greece) and urban trial sites, under conditions of vehicular traffic, network coverage, service demand, as well as considering the inherently distinct legal, business and social local aspects. Several automated mobility use cases are potential candidates to benefit from 5G such as cooperative overtake, highway lane merging, truck platooning, valet parking, urban environment driving, road user detection, vehicle remote control, see through, HD map update, media &

---

[†] Directive 95/46/EC applied until 24 May 2018. It has been repealed by Regulation (EU) 2016/679 - General Data Protection Regulation, applicable on 25 May 2018. Directive 2002/58/EC of 12 July 2002 concerning the processing of personal data and the protection of privacy in the electronic communications sector is currently under a REFIT exercise by the Commission.

entertainment. Similar to 5G-MOBIX, 5GCroCo [12] will trial 5G technologies in the cross-border corridor along France, Germany and Luxembourg. The objective is to validate advanced 5G features, such as New Radio, MEC-enabled distributed computing, Predictive QoS, Network Slicing, and improved positioning systems, all combined together, to enable innovative use cases for CCAM. In the same direction, 5G-CARMEN [13] builds a 5G-enabled corridor from Bologna, Italy to Munich, Germany to conduct cross-border trials of 5G technologies in four major use cases: cooperative manoeuvring, situation awareness, video streaming, and green driving. ICT4CART is moving on the same path with the aforementioned initiatives, with the exception that ICT4CART aims firstly to define a common architecture that can be adapted to the needs of each site and use case and additionally combines 5G with the LTE network capabilities as well as the ad-hoc capabilities of the ETSI ITS-G5 communication paradigm. Finally, two other activities that share the same objectives with ICT4CART, without the integration of 5G capabilities, are the CONCORDA project and the C-ROADS platform. The CONCORDA [14] project contributes to the preparation of European motorways for automated driving and high density truck platooning with adequate connected services and technologies, whereas the C-ROADS [15] platform is a joint initiative of European Member States and road operators for testing and implementing C-ITS services in light of cross-border harmonization and interoperability. Nationally funded projects accompany these European initiatives, e.g. the German project MEC-View [1], which works on environmental modelling on MEC servers based on infrastructure sensors in urban areas.

*1.2. Contribution and structure of the paper*

One of the first steps in ICT4CART was the development of an innovative and generic architecture that can be deployed all over Europe [16]. It supports a seamless operation with different communication technologies, i.e. ad-hoc networks (ETSI ITS-G5) and cellular networks (LTE/5G), referred to as hybrid communication, and considers multi-access edge computing (MEC) in cellular networks or similar functionality in ad-hoc networks to allow for low-latency services. It embraces these upcoming technologies and presents a unified architecture for ICT systems, thereby enabling high-value use-cases for automated driving like smart parking, intersection crossing in urban environments, lane-merging in highway scenarios and cross-border interoperability. This architecture is presented in detail in this paper, including a high-level overview as well as, in more detail, the functional view, the IT environment and data view, the communication view, and the cyber-security and privacy view.

**2. High-level architecture**

In Fig. 1, the general architecture is visualized. It shows which basic components are involved in the ICT4CART system and how they can interact with each other. The main components are the vehicles, infrastructure (sensors and processing units) and IT services (such as an OEM backend or other service providers). The basic communication technologies that are used are LTE/5G (cellular) and ITS-G5 (ad-hoc). When using 5G, slicing may be used to provide a Quality of Service (QoS) required for the kind of information to be transported, which is represented by the dashed lines in Fig. 1. There can be slices, e.g. for low-latency communication (short dashes in the figure) or high-throughput communication (lines with longer dashes in the figure). As slicing is a 5G feature, it is not available when using LTE or ITS-G5. For further details on the communication, see Section 5.

Hybrid connectivity, i.e. using ITS-G5 and LTE/5G in parallel, is also depicted in Fig. 1, as the infrastructure can provide the information via both communication channels. Also, vehicles can retrieve information over the different communication paths, either from the same source or from different sources. The vehicles in Fig. 1 are either not connected, connected with ITS-G5 (G5), with LTE/5G (5G), or with both (5G+G5). A car-to-car connection is possible via ITS-G5. The road is equipped with Road Side Units (RSUs), connected sensors, and connected traffic lights. LTE/5G base stations receive and transmit data via cellular network (green lines). ITS-G5 RSUs receive and transmit data via ITS-G5 (blue lines). Sensors and traffic lights are connected either via fibre cables (orange lines), cellular network, or via ITS-G5.

The figure also indicates the concept of Multi-Access Edge Computing (MEC) servers. MEC servers are situated close to a base station of a cellular network, providing computation closer to end devices (in the ICT4CART use cases: vehicles) and thereby avoiding time-consuming data transmission via the Internet. This concept decreases latency, especially for mission-critical data, when compared to using a cloud server anywhere in the network.

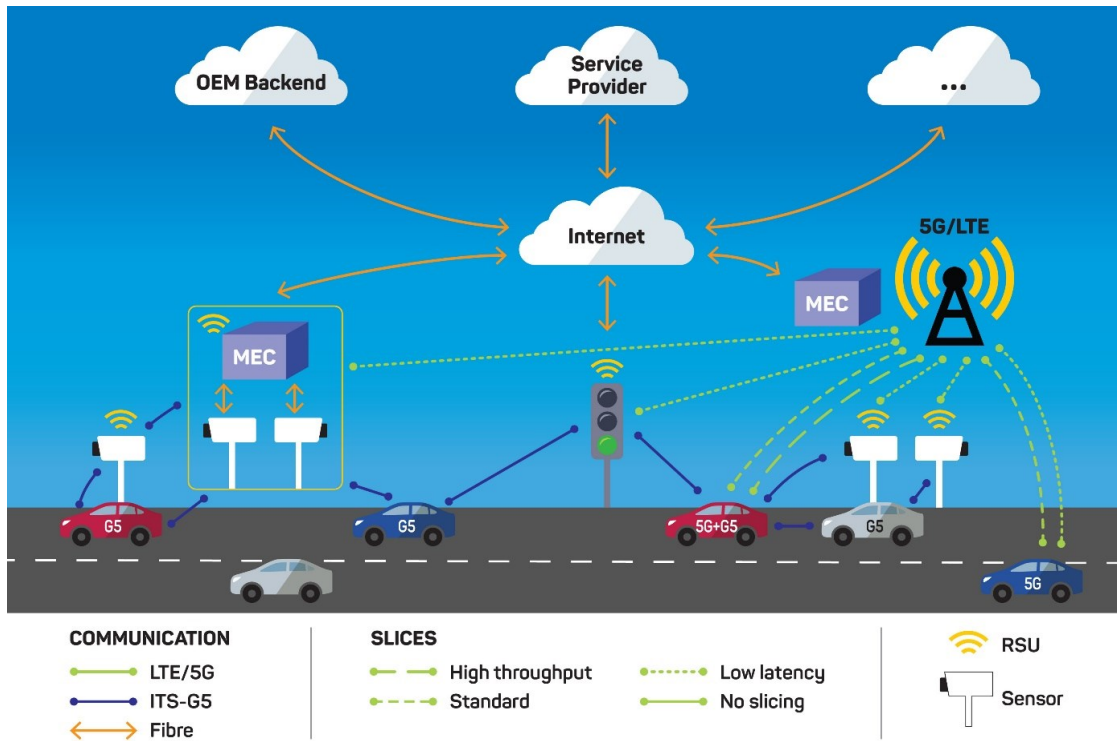

Fig. 1 ICT4CART High-Level Overview

Similar functionality like MEC servers can be provided by processing capabilities of road infrastructure like sensors in combination with RSUs. This is shown in the left part of the figure. For simplicity reasons, these processing units are also denoted as MEC here. An RSU is an ITS-G5 communication unit on infrastructure side. Such MEC servers in combination with RSUs can run ICT4CART applications or services in close proximity with lower delays than cloud services. Information from the OEM Backend, Service Providers, and other cloud services ("…" cloud in Fig. 1) can be received via the internet.

## 3. Functional view

The ICT4CART architecture consists of a collection of functional blocks. The functional view, depicted in Fig. 2, organizes the functional blocks into groups according to their common purpose: Supporting Services, Sensors and Actuators, Applications, Core Services, Hybrid Wireless Network, Cooperative Automated Vehicle (CAV), Security and Privacy. Please refer to [16] for more details on the functional view.

The **Supporting Services** group includes functional modules that are directly or indirectly related to navigation, i.e. providing information on free parking spots or charging stations via the Service Provider Gateway (SPG). The Traffic Control Centre (TCC) provides the relevant data (e.g., speed limits, location of works zones, current weather) from the road infrastructure operator. Toll Stations provide data regarding operational lanes, their occupancy rate, and available payment methods. Map Services provide an HD map with highly accurate information on traffic lanes, the locations of pedestrian crossings, etc.

The **Sensors and Actuators** group contains functions in the infrastructure to gather information about the traffic on roads as well as to manage the traffic. Traffic Sensors, like video cameras, magnetic coils, and/or LiDAR sensors, provide information regarding the current traffic in the surveyed area, e.g. to the TCC and/or directly to vehicles by an Environment Perception Model on a MEC server or RSU. Connected Traffic Lights are used to adaptively control the traffic and possibly provide information on the status.

Functions that make use of data and services from several other different groups to add business value build the **Applications** group. ICT4CART investigates some of them exemplarily: The Automation Level Road Clearance application can utilise static (e.g. legal regulations, road topology) and dynamic (e.g. traffic density, weather) information to determine a clearance for a road section and a specific automation level or the decision of a human traffic manager. The Fleet Management application keeps track of all managed vehicles and their statuses, and can make use of parking or charging station data if required for vehicles from the fleet. The Smart Parking application makes use of the parking data by assigning a parking spot and routing the requesting vehicle there.

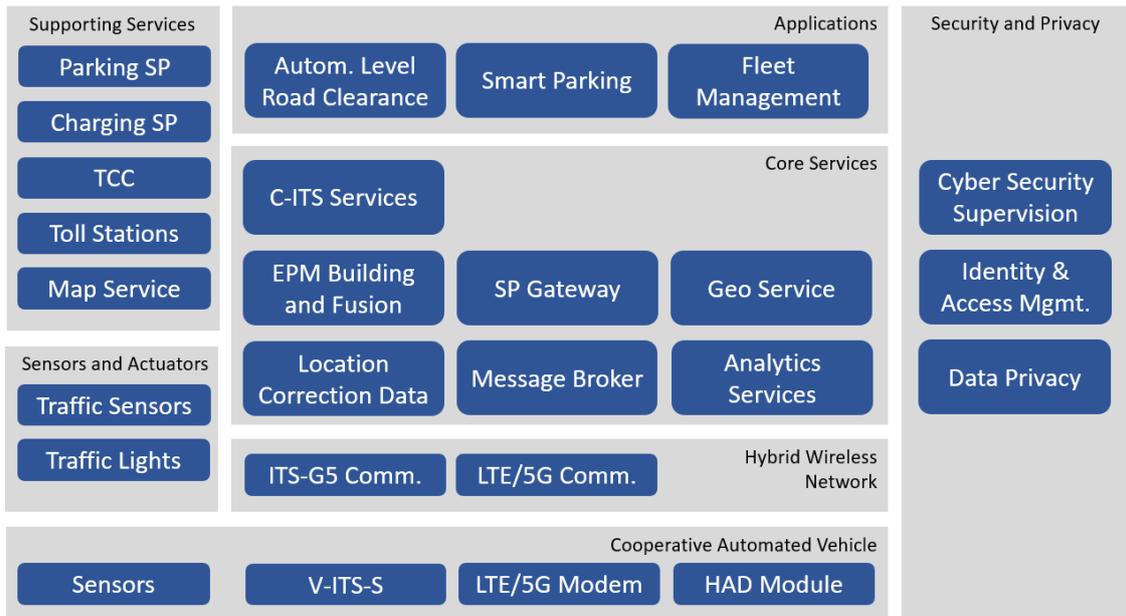

Fig. 2 Functions required for ICT4CART use cases

The **Core Services** group consists of several services that are crucial to make ICT infrastructure a key enabler for a safe and efficient deployment of HAD on European roads. This includes C-ITS Services such as Common Awareness, Decentralized Environmental Notification, Traffic Light Manoeuvre or Infrastructure-to-Vehicle Information, which are used to manage the generation, transmission and reception of C-ITS messages. Data from vehicles and/or infrastructure are received and then processed in the Environment Perception Model (EPM) Building and Fusion function, providing an EPM of the surveyed area to the road users. The Service Provider Gateway (SPG) collects the data from various different vendors and provides a single downstream interface to the communication infrastructure. The Geo Service determines the relevance radius for ITS messages, the group of wireless stations that transmit these ITS messages and the CAVs that are in the relevance radius. To allow precise positioning of the CAV, the Location Correction Data function provides correction data for Global Navigation Satellite Systems (GNSS) data by using either physical reference stations or network-based ones. The Message Broker is part of the hybrid C-ITS approach and will be used by CAVs to subscribe for messages of interest based on geographical location. There will be at least one Message Broker per country. Analytics Services refine data by using algorithms in order to prepare them for other services, e.g. making predictions for parking space availability in a specific area and time frame.

The **Hybrid Wireless Network** consists of the combination of the wireless ad-hoc ITS-G5 network and cellular networks such as LTE and 5G. This is described in detail in Section 5.

The **Cooperative Automated Vehicle** group describes the vehicle part. Vehicles equipped with both wireless communication technologies (ITS-G5 and LTE/5G) make use of both the ad-hoc information and the mobile information (hybrid communication). The on-board Sensors are a major source of information on which the CAV bases its automated driving decisions. The Vehicle ITS Station (V-ITS-S) is used for the ITS-G5 ad-hoc Wi-Fi communication with other vehicles and C-ITS infrastructure components. The LTE/5G Modem is used for the mobile connection to the radio access network (RAN) of the cellular network. The Highly Automated Driving (HAD) Module summarizes all functionality to implement automated driving using the data from its on-board sensors as well as cooperative information.

The **Security and Privacy** group contains the Identity and Access Management, Cyber Security Supervision and Data Privacy functions. Together, they ensure that all data exchanges are safe and secure while personal data privacy is preserved, as described in more detail in Section 6.

## 4. Data / IT environment view

The Data / IT environment view of the high-level architecture comprises the high-level data flows as well as the data types. This section defines the components that deal with these data flows and specifies the relevant data types. See [17] for a more detailed discussion.

*4.1. IT Environment*

A high-level component viewpoint of the ICT4CART IT environment architecture is shown in Fig. 3. As can be seen, the IT environment is distributed over two types of platforms: cloud and MEC. Services required with low latency or limited geospatial extent ($\leq$ 50 milli seconds) should be deployed on MEC servers, others should be deployed on the cloud. Vehicles should communicate with MEC services directly since MEC services and applications require low latency. However, they may communicate with cloud services using off-board access, which centralises access to these services depending on context. Please note that the vehicle is not considered as part of the IT environment. Rather, it is considered as a client. The main components of the ICT4CART IT environment architecture are introduced below:

- Off-Board Access: Acts as an intermediary between the vehicle and the external world, providing secure and privacy-preserving data exchanges. This is aligned with ISO's Extended Vehicle Concept.
- Semantic Framework: Provides reference vocabularies across the IT environment. In our current proposal, the core component of this framework is a common ontology base, which may be stored in an appropriate repository such as a triple store. A semantic server or query engine may be used to query the ontologies.
- Geo Service: Provides functionality to facilitate the handover of a vehicle from one MEC server to another as the vehicle is moving from one MEC area to another.
- Parking and Charging Station Service Gateway: Provides unified access to parking and charging station availability information, including predictions and recommendations, based on data from multiple service providers.
- Risk Maps for Automation Level Adaptation: Provide common access to context data required by the AD vehicle to assess the risk pertaining to driving in a given automation level. Data required to build the risk maps may be provided by various data sources, such as traffic control centres, MEC services, RSUs, etc.
- Collective Perception Service: Provides environment perception models (EPMs) to vehicles. Such EPMs provide information describing the physical environment (vehicles, pedestrians, objects) at a given location.
- Environmental Notification Service: Provides environmental notifications to vehicles. This component complements the information made available by the Collective Perception Service with notifications of events or situations in the surroundings that require low latency.

*4.2. Data types*

Based on the use case specifications, the following data will be exchanged throughout the ICT architecture. Data types are split into two categories: those provided by the IT environment to the vehicles, and those consumed by the IT environment. The data standards and use cases pertaining to each data type are listed in Table 1.

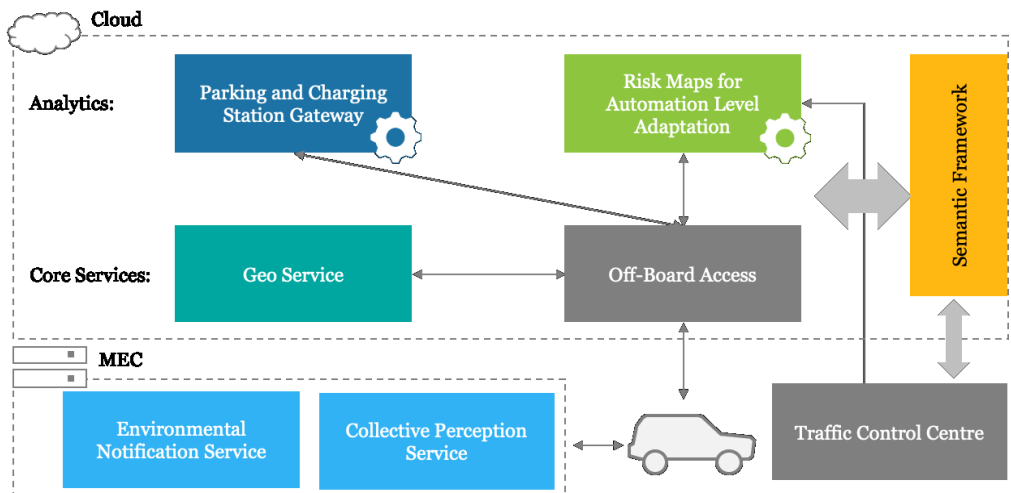

Fig. 3 ICT4CART IT Environment Architecture – Component Viewpoint

Table 1. Data Types Used in ICT4CART and Relevant Standards

| | Data Type | Relevant Standards |
|---|---|---|
| Provided by the IT Environment | Environment perception models (EPMs) for use by the automated driving (AD) vehicles | CPM as will be specified in ETSI TS 103 324 [18] |
| | Intersection map and topology | MAPEM as specified by ETSI TS 103 301 [19] |
| | Traffic light data | SPATEM as specified by ETSI TS 103 301[19] |
| | Situations and events (e.g., accident, road closure, etc.) | DENM [20], DATEX II [21] |
| | Real-time parking spot and charging station availability | DATEX II, Open Mobility Vocabulary (MobiVoc) [22] |
| | Extracted or aggregated information and predictions, such as parking predictions, traffic jams, etc. | To be defined depending on the type of output data |
| | Correction data for GNSS-based localisation | ETSI TS 103 301 [19] |
| | HD Maps | To be defined |
| Consumed by the IT Environment | EPMs, or data required to build EPMs, from RSUs and road signs and sensors | CPM, CAM as specified by ETSI EN 302 672-2 [23] |
| | Data directly received from road signs and sensors in the absence of RSUs | CPM, DENM |
| | Data received from traffic and parking services | DATEX II |
| | Data received from vehicles either directly or indirectly through RSUs | CPM, CAM |

## 5. Communication view

In the following, the Communication View of the high-level architecture is presented. After a short overview of the ITS message handling, the following subsections describe the wireless ad-hoc communication and the cellular communication. Finally, the hybrid communication based on the combination of both communication channels is sketched. For a more detailed discussion, please refer to [24].

*5.1. ITS message handling*

An ITS Station interacts with one or multiple ITS Stations to provide and/or consume transportation and infrastructure services. An ITS Station generates and transmits service-specific ITS messages, receives ITS messages from other ITS Stations, processes incoming ITS messages, and stores received, generated, and/or transmitted information. ITS messages, such as CAM, DENM, SPATEM, MAPEM, and CPM, represent user data to be exchanged between peer entities of used wireless communication technologies.

*5.2. Wireless ad-hoc communication*

Wireless ad-hoc networks such as Wi-Fi or LTE Sidelink (SL) provide decentralized communication, which allows end-user devices to exchange user data directly with each other. ITS Stations can use wireless ad-hoc communication systems to broadcast ITS messages to any other ITS Station within their proximity. An unlicensed frequency spectrum is typically used. Factors such as the used wireless ad-hoc technology and the surrounding geographical topology determine the proximity within which the ITS message can be received. The communication range is restricted to several 100m. For instance, 802.11p is used as basis for ITS-G5 in Europe.

*5.3. Cellular communication*

Cellular networks such as LTE and 5G combine a centralized communication network with a wireless access technology. The transport and connectivity services of a cellular network can be either accessed with mobiles phones (UE, User Equipment) or P-GWs (PDN Gateways). The mobile phone provides a wireless link to the communication network. P-GWs provide connectivity to external Packet Data Networks (PDNs), such as the Internet or corporate intranets.

ITS Stations using a cellular communication system typically do not broadcast ITS messages directly to other ITS Stations in their geographical proximity. Instead, an ITS Station connected to a mobile phone sends the ITS message to a central ITS Station. The central ITS Station, located in a PDN, may provide ITS message forwarding as a service. It may provide a range of additional services such as message fusing. For ITS message forwarding, the central ITS Station operates geo-location function, which stores e.g. the latest known location of ITS Stations based incoming ITS messages such as CAM. With the help of the geo-location function, the central ITS Station can determine the group of ITS Stations that are currently located in the proximity of the sender of the ITS message, and then forward this ITS message to this group of ITS Stations using the cellular network infrastructure.

Being a centralized communication network, and using licensed frequency spectra, the radio and transmission resources for an ITS Station or a group of ITS Stations are managed. This allows the mobile network operator to provide Quality of Service (QoS) between the network's edge nodes (the mobile phone and the P-GW). The key QoS attributes are latency, i.e. the transmission delay between mobile phone and PDN GW; and reliability, i.e. the packet error loss rate between mobile phone and PDN GW. If an ITS service for connected automated driving (L3/L4) requires very low total latency, the transmission delay between the ITS Station and the intermediate ITS Station needs to be minimized. This can be achieved, e.g., by operating central ITS Stations in close proximity to (vehicle) ITS Stations. A central ITS Station can e.g. be hosted on a MEC Server placed within or next to radio access network (RAN) of the cellular network, thus minimizing the number of hops through with the user data/ITS message has to be routed between the ITS-Station connected to a mobile phone and the central ITS Station.

Users and applications can have very specific requirements with respect to transport QoS and ITS service processing and storage capabilities. Network slicing allows providing dedicated virtual networks with functionalities specifically selected to meet the service or customer requirements. LTE provides rudimentary network slicing options with features such as DCN (3GPP Dedicated Core Network). Only in 5G, network slicing becomes fully available.

*5.4. Hybrid communication*

Hybrid communication deploys multiple communication networks to extend network coverage for ITS Stations and to improve ITS service availability. It allows an ITS Station to send ITS messages via one or several radio technologies, and to receive ITS messages via multiple radio technologies. The ITS Station may choose a radio technology for data transmission and reception based on QoS factors such as its reliability or the delay sensitivity of the data.

**6. Cyber-Security & Privacy View**

As in any connected contexts, cyber-attacks are a major concern. In the context of Intelligent Transport Systems (ITS), a special attention is paid on those attacks leading to privacy violations and traffic disturbance. The ICT4CART architecture proposes an approach combining secured communications compliant with the latest EU standards, as presented in [25], and monitoring of cyber-security events to make stakeholders aware of the cyber-security situation. Concretely speaking, Cyber-Security and Privacy view of the high-level architecture comprises a Cyber-Security Supervision service, an Identity and Access Management service, both including Data Privacy mechanisms. They are sketched here and detailed in [25].

*6.1. Cyber-Security Supervision service*

The role of the Cyber-Security Supervision service is to analyze cyber-security events collected from vehicles, RSUs, MEC servers and cloud services in order to detect cyber-security incidents that require to be considered by road operators, cities, carmakers, OEMs and service providers. It also assesses vulnerability against known flaws and exploits, complemented with threat knowledge. Periodic cyber-security activity and situation reports are relevant to get the big picture of the cyber-security health of a vehicle fleet, road infrastructure and parking management services for instance.

The Supervision service consists of the functionalities depicted in Fig. 4 and briefly described above. Data, as security events from embedded or network sensors or ITS messages such as CAM, DENM, etc. are first collecting real-time from data sources. Then they are processed by a rule-based correlation engine and by an anomaly detection engine based on artificial intelligence in order to detect anomalies and raise alerts. These one could be directly reported in dashboards and reports, or could be also further investigate by cyber-security experts, who can fill in incidents reports by gathering relevant information. Some countermeasures could be put in place through the reaction enforcement capacity to mitigate a detected incident. Linkability is a Data Privacy mechanism, described more deeply in Section 6.3.

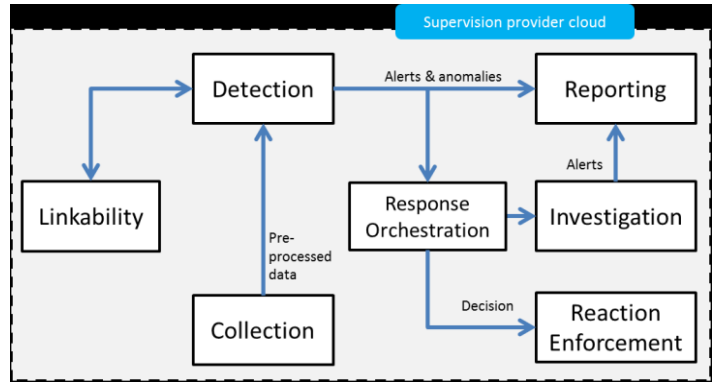

Fig. 4 Functional architecture of the cyber-security supervision service

*6.2. Identity and Access Management Service*

The role of the Identity and Access Management (IAM) service is to provide authentication and authorization functionalities to ITS-S and Users in compliance with European standards, and integrates innovative mechanisms to grant respect for privacy.

The IAM service described through consists of the following functionalities:
- Public Key Infrastructure (PKI): The PKI's architecture proposed integrates Root Certificate Authority, Enrolment Authority, that delivers enrolment credential to authenticate and grant access to ITS communications; and Authorization Authority, that delivers authorization tickets to authorize the use of specific ITS services.
- Users Access module: This module includes the authentication module that receives identity information validates them and generates access token; and the authorization module that offers access control and web request filtering capabilities based on rights contained in the token.
- IAM Directory: Is the identities and rights referential. All users, ITS-S and their rights are stored into this directory.
- Administration Application: Provides an application to manage users, devices and right models. From this application, an administrator can activate or deactivate a device, and revoke a certificate.
- Linkability Manager: This module ensures privacy and confidentiality for ITS-S, is detailed in Section 6.3.

*6.3. Data privacy mechanism*

Linkability is the Authority issuing pseudonymous credentials for users and devices to be able to issue pseudonymous signatures that can be linked. For example, in the case of the Supervision service, linkability is used within the event correlation process to detect collected events referring to the same vehicle, even if the vehicle identification has been pseudonymised.

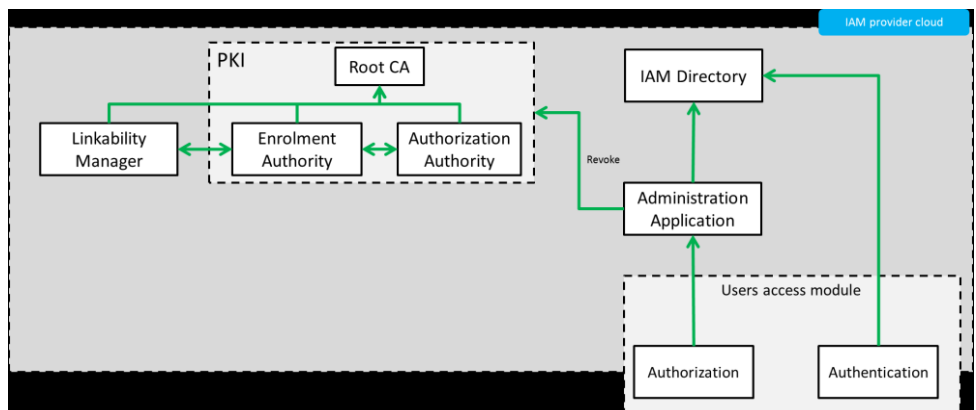

Fig. 5 Functional architecture of the IAM service

## 7. Conclusion

In this paper, a high-level architecture was presented, showing the basic components that are involved in the ICT4CART system and how they can interact with each other. This was followed by a functional view on the architecture, which explained the functional components in more details. The outlined communication architecture incorporates upcoming technologies such as MEC, ITS-G5, 5G and network slicing. We also introduced new services to address challenges in Cyber-Security, Data Privacy and Communication, e.g., the Identity and Access Management Service and the Supervision Service. The aim of this architecture is to cover all of the different ICT4CART use cases and to offer the ability for deployment in all test sites and beyond by being generic enough to be deployed for automated driving in general and to allow more use cases to be integrated.

**Acknowledgements**

The ICT4CART project has received funding from the European Union's Horizon 2020 research & innovation programme under grant agreement No. 768953. The content reflects only the authors' view and the European Commission is not responsible for any use that may be made of the information it contains.